\documentclass{IEEEtran}

\usepackage[colorlinks,urlcolor=blue,linkcolor=blue,citecolor=blue]{hyperref}
\usepackage[nolist]{acronym}
\usepackage{graphicx}

\begin{document}


\title{Towards Sustainability in 6G and beyond: Challenges and Opportunities of Open RAN}

\author{\IEEEauthorblockN{Hamed Ahmadi \IEEEmembership{Senior Member,~IEEE}, Mostafa Rahmani \IEEEmembership{Member,~IEEE}, Swarna Bindu Chetty \IEEEmembership{Member,~IEEE}, Eirini Eleni Tsiropoulou \IEEEmembership{Senior Member,~IEEE}, Huseyin Arslan \IEEEmembership{Fellow,~IEEE}, Merouane Debbah \IEEEmembership{Fellow,~IEEE}, Tony Quek \IEEEmembership{Fellow,~IEEE}
\thanks{H. Ahmadi, M Rahmani and S.B. Chetty are with the University of York, UK. E. E. Tsiropoulou is with University of Arizona, USA. H. Arslan is with Istanbul Medipol University, Turkiye. M. Debbah  is with Khalifa University, UAE. T. Quek is with Singapore University of Technology and Design, Singapore.}
}
}

\maketitle

\enlargethispage{10pt}

\begin{abstract}
The transition to 6G is expected to bring significant advancements, including much higher data rates, enhanced reliability and ultra-low latency compared to previous generations. Although 6G is anticipated to be 100 times more energy-efficient, this increased efficiency does not necessarily mean reduced energy consumption or enhanced sustainability. Network sustainability encompasses a broader scope, integrating business viability, environmental sustainability, and social responsibility. This paper explores the sustainability requirements for 6G and proposes Open RAN as a key architectural solution. By enabling network diversification, fostering open and continuous innovation, and integrating AI/ML, Open RAN can promote sustainability in 6G. The paper identifies high energy consumption and e-waste generation as critical sustainability challenges and discusses how Open RAN can address these issues through softwarisation, edge computing, and AI integration.
\end{abstract}


\section{Introduction}

{T}he discussion around \ac{6G}, similar to previous generation, has been initiated with promises on improvement of conventional \acp{KPI}, such as peak data rate and latency \cite{Ahmadi_6G_DT21}.  
It also aims for significantly improved values for some more modern \acp{KPI} like reliability, high-precision positioning, device intensity, spectral efficiency, and energy efficiency.

\begin{figure*}[h]
    \centering
    \includegraphics[width=.9\textwidth]{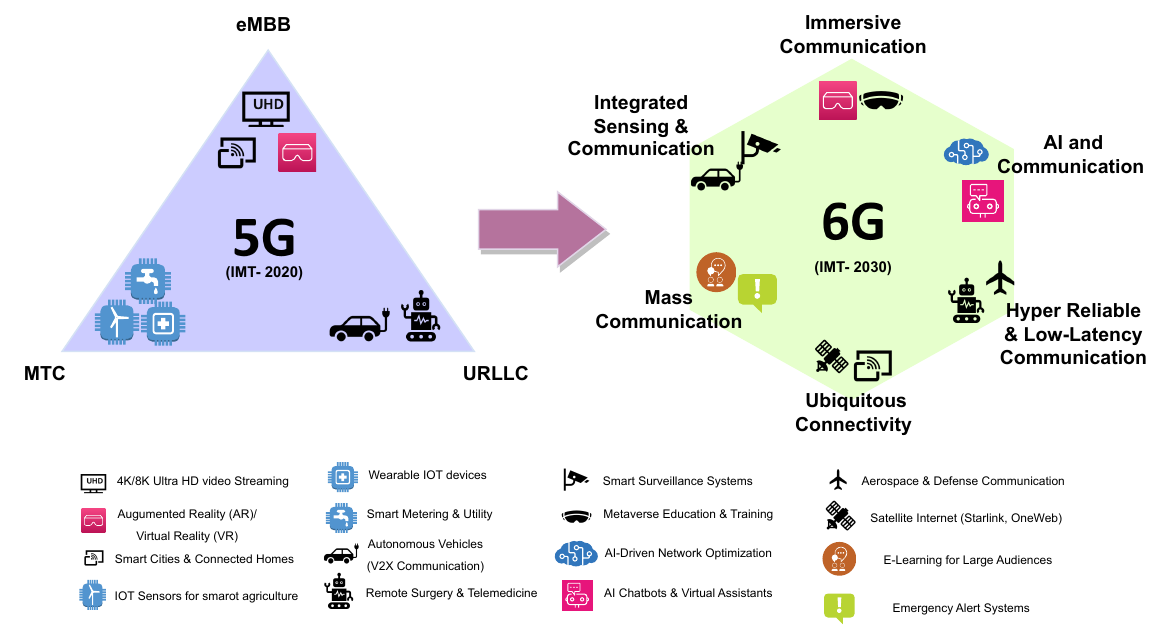}
    \caption{5G application scenarios vs 6G.}
    \label{fig:5G-6G}
\end{figure*}

The famous \ac{5G} triangle of \ac{eMBB}, \ac{MTC} and \ac{URLLC} in \ac{6G} is expected to be updated by the
hexagon of immersive communication, massive communication, hyper reliable and low-latency communication, ubiquitous connectivity, integrated sensing and communication, and \ac{AI} and communications.
These six categories of services better enable us to adjust and segregate the requirements of new applications, such as advanced autonomous systems, immersive \ac{XR}, pervasive and ubiquitous \ac{IoT}, \ac{DT}, and \ac{AI}-native services. 
The IMT-2030 defined usage scenarios have four design principles of sustainability, connecting the unconnected, ubiquitous intelligence, and security-and-resilience which need to be considered in all usage scenarios \cite{IMT2030}.
Figure \ref{fig:5G-6G} presents this evolution.
%

Key technologies driving the development of \ac{6G}, as outlined in the IMT-2030 framework, encompass a range of emerging technological trends. These include the integration of \ac{AI} and \ac{ML} for smarter network operations and autonomous management, \ac{ISAC} for enhanced data collection and transmission efficiency, and the use of higher frequency bands in the \ac{THz} range for increased data throughput. Additionally, channel optimisation through \ac{RIS} and holographic \ac{MIMO} surfaces is poised to significantly enhance signal propagation and network capacity. Other pioneering technologies, such as \ac{OAM} multiplexing, and Orthogonal Time Frequency Space (OTFS) are considered as  contending waveform for \ac{6G} \cite{Ahmadi_6G_DT21}. However, some of these technologies are still in their infancy stage and might have not achieved the expected success.
\ac{6G} is also set to enhance security through technologies such as quantum cryptography, ensuring robust protection in an increasingly connected world \cite{6G_testbed_wang23}.

Delivery of the promised services and \acp{KPI} of \ac{6G} should be in a resource efficient manner. Similarly pricing, social aspects and public acceptability must also be considered in \ac{6G}. To achieve these goals, network softwarisation and cloudification activities, which has started in \ac{5G}, should be continued in \ac{6G} with a much higher speed and with adoption of recent innovations. Softwarisation and cloudification of networks enable virtualisation and independence of network functions from specific hardware. This move leads to more robustness and efficiency in the network. Additionally it facilitate integration of \ac{AI}/\ac{ML} for a more intelligent network automation. All of these can lead to reduction of energy consumption, \ac{CAPEX} and \ac{OPEX}. 

One approach towards softwarisation/cloudification of the \ac{RAN} is the Open \ac{RAN} technology and its reference architectures/recommendations proposed by the O-RAN Alliance known as O-RAN \cite{Polese24_ORAN_6G}.  This approach paves the way for open, interoperable, observable, and intelligent cellular networks, allowing multiple vendors to contribute to a more flexible and efficient network architecture. 
Crucial to this evolution is \ac{AI}/\ac{ML}, which provides the tools needed for self-organising O-RAN systems. By integrating \ac{ML}, O-RAN enhances network performance, meeting the complex demands of \ac{6G}. The integration of intelligent and optimised management techniques is supported by \acp{RIC}, empowering the network to dynamically adapt to changing conditions, optimise resource allocation, and enhance overall performance. Furthermore, the adoption of open interfaces enables disaggregation of \ac{RAN} components, fosters interoperability and allows for the integration of diverse components from multiple vendors, promoting innovation and avoiding vendor lock-ins. This collaborative framework fosters rapid technological advancements and cost-effective deployments, driving the future of wireless communication.    

Many researchers associate telecommunications sustainability primarily with \ac{EE}, yet they often overlook that it extends beyond mere \ac{EE}. True sustainability in telecommunications encompasses a broader commitment to environmental stewardship. This includes minimising \ac{e-waste} through the design of durable, upgradable, and recyclable hardware, and transitioning to renewable energy sources like solar and wind to power network operations. Such integrated strategies ensure that technological advancements do not compromise environmental health. O-RAN promotes the use of standardised, off-the-shelf hardware, which can be easily repurposed and upgraded. This reduces the need for constant re-deployment of proprietary hardware, cutting down on \ac{e-waste}.

This paper explores how the O-RAN plays a crucial role in advancing the sustainability of \ac{6G}. Through its open interfaces and standardised protocols, O-RAN encourages a competitive and diverse supplier ecosystem, reducing reliance on single vendors and promoting the adoption of green technologies through innovation and network democratisation. This not only enhances network scalability and efficiency but also aligns with sustainable practices, optimising energy usage and reducing \ac{e-waste}. By examining the implementation of O-RAN, this study highlights how innovative network management can contribute to a more sustainable and environmentally friendly telecommunications landscape in the era of 6G.
In this paper we firstly review the concept of sustainability beyond \ac{EE}, then we investigate how \ac{6G} can become sustainable. Later we explain how O-RAN can contribute toward sustainability of \ac{6G}. Finally, before the conclusion we present a case study on the use of O-RAN to enable sustainability of \ac{CF} massive MIMO.

\section{The path towards sustainability}
The United Nations World Commission on Environment and Development defines sustainable development as ``development that meets the needs of the present without compromising the ability of future generations to meet their own needs". The authors of \cite{Giovannoni2013} identified three main concepts that cover the ubiquity of the notion of sustainability. These concepts are business/economic sustainability, environmental management and social responsibility. In the context of mobile networks and \ac{6G}, similar to other domains, these notions are interconnected. In the rest of this section we will clarify each of the definitions of sustainability in the context of \ac{6G}.

\subsection{Business/Economic Sustainability}  
As defined in \cite{Giovannoni2013} business sustainability is the
business of staying in business or in other words, meeting the needs of direct and indirect stakeholders of the company/sector without compromising its ability to meet the needs of future stakeholders as well. From the network operators' point of view generating revenue depends on many factors that include the types of service, scalability, affordability, etc. The mentioned use cases and applications of \ac{6G} have the potential to create a reasonable market demand for \ac{6G}. However, their market size and demand level depend on other sectors like entertainment, robotics, electronics, battery and manufacturing. Studying these in detail is beyond the scope of this work. Let us here focus on the expenses that also relate to environmental and social sustainability.

\ac{6G} aims to significantly improve \ac{EE}, with the \ac{KPI} expected to increase by $100$ times, from $10^7$  bits/J in 5G to $10^9$ bits/J in \ac{6G}, a crucial aspect as the sector moves towards sustainable telecommunications \cite{6G_testbed_wang23}. However, it is critical to understand that enhanced \ac{EE} does not necessarily translate to reduced overall energy consumption. While per-unit data transmission becomes more efficient, the aggregate energy consumption might still rise due to the exponential increase in data demand. This can translate into energy cost increase in \ac{6G} leading to potentially higher cost of the final service. 

To deliver the new services a significant hardware and network equipment upgrade is required. This upgrade is costly and time consuming in terms of procurement procedure, installation and retraining the operations team. Removal and recycling of the replaced infrastructure also imposes extra cost to the network operators. Thus, business sustainability in \ac{6G} requires managing these costs in an environmentally and socially sustainable way. 

\subsection{Environmental sustainability}
Environmental management includes consideration of a wide range of indicators including not only energy efficiency, but also material
efficiency, biodiversity, emissions, water consumption, and waste. \ac{ICT} has significantly contributed in different forms to sustainable growth like supply chain efficiency, manufacturing, or (more recently) reduction of unnecessary travel by facilitating working from home. However, it also significantly contributes to world's total energy consumption, emissions and \ac{e-waste}. \ac{6G} will introduce several new services that require new higher data rates, lower latency and wider coverage. All of these as mentioned before, mean that more energy will be consumed. While energy consumption has been a measure for development for a long while, it is also closely related to greenhouse gasses release and potentially climate change. Therefore, energy efficiency is a key factor in environmental sustainability especially in the the case of \ac{6G}. 

 E-waste in the telecommunications sector has become a significant environmental challenge, due to the industry’s rapid growth and the swift obsolescence of its technologies. As the telecommunications sector transitions towards more advanced 6G, the potential for \ac{e-waste} increases significantly. Like earlier generations, 6G equipment will contain hazardous materials that pose risks to environmental health if not properly managed. Also, the continuous cycle of replacing old equipment with new not only consumes valuable resources but also results in significant loss of precious materials if recycling is not efficiently executed. True environmental sustainability in \ac{6G} cannot be achieved unless we also address the energy consumption/efficiency and \ac{e-waste} issues. 

\subsection{Social responsibility}
Traditionally social responsibility is
defined as the willingness and the approach towards using the resources towards broad social
ends and not simply for the narrowly circumscribed interests of private persons and firms \cite{Giovannoni2013}. 
However, more in modern literature like UN Sustainable Development Solutions Network report \cite{UN2024} social responsibility refers to avoiding the exhaustion of valuable and irreplaceable resources which denies them from future generations to gain short-term development/advantage. While \ac{ICT} and telecommunication significantly contribute to several social development aspects particularly in terms of accessibility of youth and future generations to education and healthcare, excessive use of materials for non-recyclable \ac{ICT} equipment and fossil fuels for electricity production are in contradiction with the modern view of social responsibility.

\section{How to make 6G and beyond sustainable solutions?}
In all three sustainability key factors that we discussed in previous section, challenges in energy consumption and generation of e-waste are the most immediately relevant to \ac{ICT} and particularly \ac{6G} and beyond.

\subsection{E-waste reduction via softwarisation}
A path towards satisfying all three main concepts of sustainability is reduction of the e-waste. Softwarisation and/or decoupling the software from its running platform is a major step towards this goal. Running network functions in specialised hardware increases the performance and efficiency of the process. However, with every service upgrade there might be a need for replacing the equipment with newer ones. Virtualisation and softwarisation of network functions enable using \acp{GPP} and hardware in the network which will operate for a longer duration with software updates and minimal hardware upgrades.
For example a softwarised network function that is detached from any particular hardware  can be executed on a \ac{GPP}. However, the same
function, if not softwarised requires particular hardware and is attached to it. Upgrading that network function requires the hardware upgrade i.e. hardware change while softwarised network functions can be upgraded without hardware upgrade and if more processing or memory is required it can be added to the general purpose computers or servers that they are running on. 
As we move away from usage of specialised equipment (which are needed in many cases like the transceivers) towards more \acp{GPP} the possibility of the equipment redeployment in another part of the network increases. Currently most of the advancements in virtualisation and softwarisation have been at the core and very little has been done at \ac{RAN}. Thus, \ac{6G} and beyond should retain and improve the softwarisation advancements of \ac{5G} core and expand the softwarisation to \ac{RAN} as well to be able to reduce core and \ac{RAN} e-waste. 


\subsection{Sustainable energy consumption}
The telecommunications industry has become and will remain a major consumer of energy, driven primarily by data centers, network operations centers, and a comprehensive array of transmitting equipment. Finding ways to reduce or manage \ac{6G}'s energy consumption and using renewable energy sources are two main approaches to sustainable energy consumption \cite{Liang_EC_ORAN24}.

\subsubsection{Renewable energy at the edge}
To effectively integrate renewable energy into the telecommunications, several innovative strategies are being implemented, transforming the industry's energy consumption patterns. This is enabled by architectural improvements in \ac{6G} that improved integration of edge computing in the network. Edge computing and cloud-edge continuum are being considered in the architecture of \ac{6G} and O-RAN which has not been done in the previous generations \cite{Zhao_edgeComputing_21}. For example, in O-RAN architecture the near realtime \ac{RIC} which handle many \ac{AI} related processing can be placed at the edge.

Traditionally, the base stations in the remote or rural area are relied heavily on diesel generators to maintain their operations. Now, solar panels are being installed directly at these sites. These panels harness the solar energy to power the equipment and the essential cooling systems, drastically reducing the reliance on fossil fuels. In areas with favorable wind conditions, operators are deploying to wind turbines. We can now easily consider small to medium-scale turbines spinning gracefully, capturing the wind's energy to power network equipment. Advancements in edge computing and battery technology that reduced the required space and cost, and improved the efficiency, facilitated the usage of renewable energy in Telecommunications. The advanced batteries can store excess energy generated during peak energy production periods, ensuring a steady and reliable power supply even when renewable sources are temporarily unavailable. Additionally, many computational tasks can be handled at the edge when there is excess of generated green energy and during the peak consumption time, if possible and not violating \ac{QoS} requirements, they will be loaded to the cloud. It also should be noted that deployment of renewable energy sources and energy storage at the edge might not be always feasible and/or economically viable as the space and equipment requirements and cost varies in different locations and countries.
Edge computing and cloud-edge continuum are being considered in the architecture of \ac{6G} and O-RAN which has not been done in the previous generations. For example, in O-RAN architecture the near realtime \ac{RIC} which handle many \ac{AI} related processing can be placed at the edge.
There also have been significant works in using green energy in cloud and data centres. 

\subsubsection{Energy sufficiency}
As defined in \cite{darby2018energy} ``Energy sufficiency is a state in which people’s basic needs for energy services are met equitably and ecological limits are respected". The concepts like ``people's basic needs" are place for debate and there exist disagreements around them. Defining the basic needs in the context of \ac{6G} will be debatable as well. By defining the basic needs as the minimum requirements of the service to operate we will reach our existing energy efficiency goals, while energy sufficiency also relies more on cultural and life style changes.  It can be considered as similar to cycling or using public transport to work instead of driving. Achieving energy sufficiency goals depend on user side behavioural change. This can be as simple as not uploading every document/file on the cloud, instead carrying a USB memory stick, or not streaming \ac{UHD} media on mobile phones where the difference in quality will not be significant on the small screen. Similar actions can be considered in \ac{MTC} and \ac{URLLC} where the end users are encouraged to consider lower energy consuming solutions. 

Energy sufficiency in networks require definition of actionable guidelines for operators and users which needs to be investigated in detail by technologists, policy makers and even social scientists. These actions may include dynamic cell coverage and capacity balancing during different hours of a day on the operators side, reducing the \ac{UHD} streaming/downloading media on the user side, and encouraging shared community owned neutral host access points/base stations on the community side.  
Currently, this research is at infancy level and there is no agreed actionable guidelines.

\subsection{AI/ML and sustainability}
After the discussion about sustainable approaches for e-waste and energy consumption reduction, it is time to talk about the elephant in the room, \ac{AI}. \ac{6G} is expected to be the \ac{AI}/\ac{ML} native network technology\footnote{An \ac{AI} native technology is a technology in which \ac{AI} is a core part of its design, deployment and operation.}. \ac{AI}/\ac{ML} will have a significant role in main \ac{6G} technologies from core to \ac{RAN} \cite{Liang_EC_ORAN24}. It will include intelligent resource allocation from spectrum level to network planning including new technologies like \ac{ISAC} and \ac{RIS}. Also \ac{AI}/\ac{ML} will be heavily used in important usecases like Autonomous systems, \ac{CAV}, \ac{UAV}, \ac{XR}, and \ac{DT} \cite{Ahmadi_6G_DT21}. While \ac{AI}/\ac{ML} solve complex problems that could not be solved with conventional methods, its power hungry nature created challenges of energy consumption and e-waste. Recently \ac{LLM} and Generative \ac{AI} become very popular in different domains including telecommunications. These powerful \ac{AI} tools can be used for network planning, join source and channel coding, semantic communications, and many other \ac{6G} and beyond technologies \cite{Bariah_GenAI24}. However, these massive benefits come with a huge cost. \acp{LLM} and Generative \ac{AI} consume enormous amount of power to train. This amount of consumed energy also leads to a huge $CO_2$ emissions; as presented in \cite{bender2021dangers} training a big transformer model causes more $CO_2$ emission than $55$ average person per year. The authors of also \cite{bender2021dangers} highlighted that ``training a single BERT base model (without hyperparameter tuning) on GPUs is estimated to require as much energy as a trans-American flight" which simply reminds us the importance of considering energy sufficiency concept, that aims to reduce the total energy consumption, in modern \ac{AI} in addition to energy efficiency. 

Reduction of \ac{AI} training time and energy is an active research topic and several methods have been discussed in the literature \cite{Liang_EC_ORAN24}. Methods like transfer learning update a trained model in a base environment with inputs from target environment to reduce the training time and energy. Federated learning, aggregates the collected models from distributed agents, and then feeds back the aggregated model to the agents. This approach enables learning at the edge and reduces the overhead of transmitting the raw data which also preserves the privacy. For \ac{LLM} fine tuning the a pre-trained model will significantly save energy and cost compared to retaining.

In addition to high power consumption, training advance \ac{AI} models rely on large quantities of data, and significant processing and storage capabilities. Large data collection, storage and processing all demand new electronics contributing to the increase of e-waste. These highlight the importance of efficient and ethical use of \ac{AI} and presents that \ac{AI}-based approaches are not necessarily addressing the sustainability challenges. 


\section{Open RAN: a recipe to 6G and beyond sustainability}

\begin{figure}
    \centering
\includegraphics[width= 0.9\columnwidth]{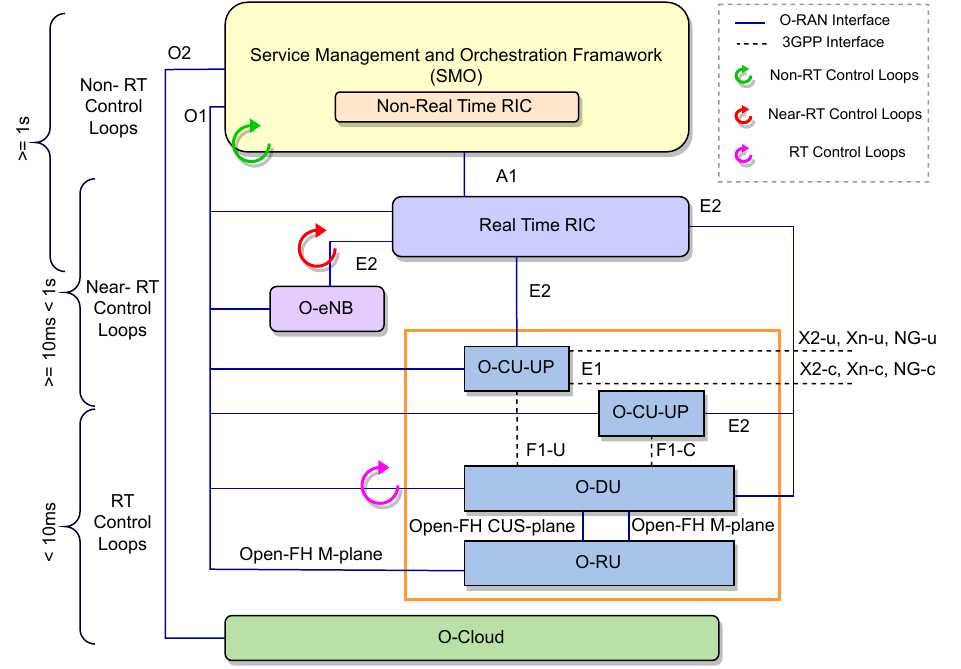}
    \caption{Architecture and main components of O-RAN framework}
    \label{fig:oran-archi}
\end{figure}

\subsection{Open RAN}
Cellular networks are undergoing a major transformation, primarily driven by the Open networking and Open RAN paradigms.
The Open \ac{RAN} paradigm focuses on software-based solutions, disaggregation of network components, open interfaces, and the use of programmable ``white-box" hardware to replace traditional, closed architectures. Open networking covers open standards across various network elements, including routing, switching, and core.
The goal is to create more flexible, multi-vendor, data-driven networks optimised for performance and adaptability \cite{Polese24_ORAN_6G}.

This change is spearheaded by the O-RAN Alliance, which consists of
network operators, vendors, and academic partners. The alliance aims to standardise the Open \ac{RAN} architecture, its components, and open interfaces to ensure interoperability across different vendors. This standardisation also supports real-time monitoring, data collection, and cloud integration. For instance by implementing the 7.2x functional split \cite{Polese24_ORAN_6G}, the O-RAN framework builds on the disaggregated architecture of the \ac{3GPP} \ac{gNB}, dividing functionalities  \acp{CU}, \acp{DU}, and \acp{RU}. The \ac{RIC} is introduced to manage both near-real-time (near-RT) and non-real-time (non-RT) \ac{RAN} control through specialised software applications called xApps and rApps. These elements expose adjustable parameters, functionalities, and data streams, such as \acp{KPM}, allowing xApps and rApps to optimise network behavior based on operator goals and current conditions using advanced \ac{AI}/\ac{ML} algorithms \cite{Liang_EC_ORAN24}. Figure \ref{fig:oran-archi} presents the architecture and main components of O-RAN framework.


\subsection{Open innovation for RAN}
Open innovation is an approach towards innovation in which an organisation does not only rely on its own in-house build ideas and knowledge and is open to share information and knowledge. There are many benefits associated with open innovation including lower innovation costs, shorter time to market, higher innovativeness, and higher number of innovations \cite{HUIZINGH_innovation2011}. By lowering the innovation cost and shortening time to market open innovation enables the smaller firms to survive in a competitive market leading to democratisation of the supply chain. O-RAN is an approach to incorporate open innovation in the telecommunications landscape which traditionally have been dominated by an oligopoly of limited number of vendors. 

Disaggregation and open interfaces in O-RAN allow smaller equipment vendors to enter the market by providing specific and specialised \ac{RAN} components e.g. \ac{RU}. O-RAN has made software innovation and integration of \ac{AI} in \ac{RAN} significantly easier by introducing \ac{RIC}. Within O-RAN framework, smaller software companies can develop xApps/rApps to enhance network performance in different domains. These xApps/rApps can make self-organising networks are become a reality, and are capable of autonomously adapting to changes in channel conditions, network states, and traffic patterns. 

\begin{figure*}[h]
    \centering
\includegraphics[trim={2cm 2cm 2cm 2cm}, clip, width=1\linewidth]{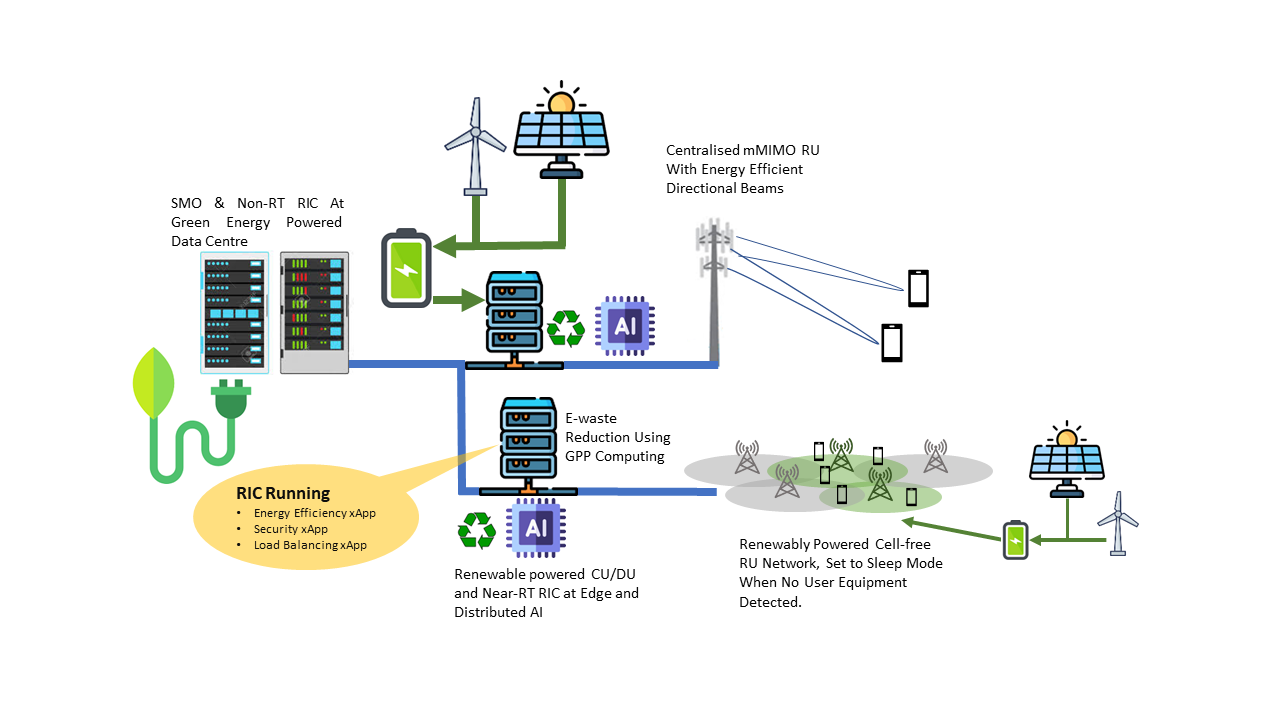}
    \caption{Sustainable O-RAN based 6G network.}
    \label{fig:o-ran-sustain}
    \vspace{-.5cm}
\end{figure*}

\subsection{O-RAN and sustainability concepts}

As argued in \cite{HUIZINGH_innovation2011} open innovation is a well established approach to business sustainability in technological domains and firms. In terms of environmental sustainability, open innovation enables continuous innovation which supports smaller improvements in performance, efficiency and/or user friendliness. Similarly, user feedback on environmental issues, social responsibility, can reach the production stage much faster in ecosystems that embrace open and continuous innovation. O-RAN creates an ecosystem in which open and continuous innovation for \ac{6G} and beyond is supported \footnote{It must be noted by open innovation we do not mean open source as in O-RAN the open innovation is being implemented in both close and open source. The open interfaces and standardised point to plug in third party components like xApp/rApps enabled open innovation at O-RAN.}.Using this approach all the aforementioned approached to make \ac{6G} sustainable can be implemented. 

\subsubsection{O-RAN and E-waste} Standard and open interfaces and softwarisation made O-RAN a promising solution for e-waste generation in \ac{RAN}. Via softwarisation several components like \ac{CU} and \ac{DU} can be deployed as software packages on general purpose hardware which reduces the requirement of \acp{gNB} for specialised hardware which become obsolete after few years. O-RAN open interfaces facilitated interoperability and let network components from one vendor to work with components from other vendors. This enables network equipment to be reused in other part of the network when they are not efficient enough for other parts anymore. For example if an operator decides to use newly developed multi-band \acp{RU} in demand hotspots like city centres, it can reuse the existing ones in parts of the network with lower load like suburbs or rural areas without fear of these \acp{RU} not being compatible with other equipment used there. It should also be noted that there will be an increased level of hardware deployments required by disaggregated O-RAN components which have their own additional environmental impact and generation of e-waste.

\subsubsection{O-RAN and energy consumption} 
Edge computing on general processors has been made possible in O-RAN architecture via disaggregation. In O-RAN \ac{CU}, \ac{DU}, and \ac{RIC} can be deployed in separated units at the edge of the network. Each of these can benefit from renewable energy sources deployed nearby them like small/medium size solar panels. \ac{RIC} in O-RAN architecture is the host for most of the software-related innovation. It allows (verified) third-party software as xApps/rApps to run and make decisions on controlling \ac{RAN} parameters. In \cite{Liang_EC_ORAN24}, we presented an \ac{ML}-based xApp that makes energy saving for the \ac{RAN} by handing over the \acp{UE} of underutilised \acp{RU} to its neighbouring cells and putting the \acp{RU} to sleep mode. xApps/rApps are also capable of managing \ac{RU}'s power, load and \ac{UE} admission. However, these controls are managed differently on \acp{RU} manufactured by different vendors. Therefore, different xApps, provided by different vendors, for the same purpose can be used in different parts of the network. This level of flexibility and diversity, if not impossible, is very hard to be delivered in rigid not open network architectures. Additionally as new hardware e.g. \ac{RU} enter the market, through open innovation xApp development will be able to catch up and release updated versions with enhanced energy saving features. 

While all mentioned about the energy consumption reduction approaches in O-RAN there are still debates whether O-RAN, due to its reliance on increased processing power and more complex management systems, might actually increase overall energy consumption rather than reduce it. This debate is a valid one while end-to-end energy consumption results are missing and the network energy consumption very much depends on the hardware as well. However, our simulations in \cite{Liang_EC_ORAN24} presents than on RAN side O-RAN outperforms LTE e-NodeB on low and medium transmit power. A more detailed O-RAN power consumption analysis considering RU/DU/CU distance and disaggregation level is provided in \cite{Baldini_EC2024}.

\subsubsection{O-RAN and AI} In O-RAN architecture \ac{AI}/\ac{ML} is a core component. The architecture introduced \ac{RIC} to be the host for \ac{AI}. As explained in \cite{Liang_EC_ORAN24}, O-RAN Alliance recommended different \ac{AI} deployment scenarios leaving multiple choices to developers and creating room for innovation in \ac{AI} deployment in O-RAN. The O-RAN Alliance recommendation also allows implementation of distributed and federated learning in which the near-realtime \acp{RIC} at or near to the edge are learning the local model and non-realtime \ac{RIC} hosts the aggregator algorithm. \ac{AI}-based xApp/rApps can significantly contribute to making O-RAN networks more sustaibable by optimising energy consumption, balancing the O-RAN network components, \ac{RU}/\ac{DU}, 
 activate time and power consumption depending on their renewable energy  availability, scheduling and predictive maintenance, etc. In Figure \ref{fig:o-ran-sustain}, we present a sustainable O-RAN based \ac{6G} network, summarising our discussions in this section.

In addition to all the benefits of \ac{AI}/\ac{ML}, as mentioned before there are also potential downside of them such as the environmental impact of data centers and the energy cost of AI processing which should not be ignored and they will remain open issues. Given these considerations, it is important to systematically study and quantify the impacts versus the enablement provided by \ac{AI}/\ac{ML}, particularly in distributed applications such as those in O-RAN architectures. This will help ascertain whether the benefits can indeed outweigh the drawbacks.





\section{Case Study - Cell Free Massive MIMO and O-RAN}

In the context of advancing network technologies towards \ac{6G}, \ac{CF} architecture represents a transformative shift from traditional cellular designs. Unlike centralized systems that rely on cell towers to serve defined areas, \ac{CF} architecture employs a dispersed array of multiple \acp{AP} throughout the coverage area. This strategic distribution not only ensures uniform signal quality across all users, thereby reducing the network’s susceptibility to environmental factors like shadowing and fading, but also marks a shift towards more resilient and adaptable network infrastructures. 
This architecture is not just about spreading out the hardware; it's a smarter approach to managing network resources. By allocating network resources based on real-time demand and location data, \ac{CF} architecture optimizes system performance with minimal energy waste. Such efficiency translates into significantly lower overall power consumption, addressing key sustainability challenges faced by modern telecommunication systems \cite{CF_MMIMO_2022}. 
Furthermore, the dynamic resource allocation inherent in \ac{CF} architecture necessitates an intelligent and agile approach, which is enabled by \ac{AI}-based xApps/rApps within the O-RAN framework. This integration showcases how O-RAN not only supports but enhances the capabilities of \ac{CF} architectures, making a compelling case for its role in the sustainable evolution of communication systems.
Thus, O-RAN emerges not just as a facilitator but as a critical enabler of \ac{CF} architecture, providing the necessary tools and open interfaces that allow for such sophisticated deployment and management. The application of O-RAN in \ac{CF} architectures exemplifies how technological innovation can be harnessed to meet stringent energy efficiency and sustainability goals in preparation for \ac{6G} networks.

While \ac{AI} system inherently require significant computational power and thus energy, the \ac{CF} architecture significantly aids in offsetting these demands. The dispersion of processing across multiple \acp{AP} not only enhances signal quality and network resilience but also allows \ac{AI}-driven optimizations to be executed more efficiently at local nodes. By implementing AI solutions that are optimized for energy efficiency and tailored specifically to CF architectures, we can maximize the benefits of dynamic resource management, ultimately leading to lower overall power consumption and a more sustainable network infrastructure.

In the following, we examine the performance and power consumption of O-RAN CF-mMIMO technology compared to traditional small-cell systems within a virtualised architecture \cite{demir2024cell}. We also confirm how the end-to-end orchestration scheme aims to reduce overall power consumption by jointly managing radio, optical fronthaul, and cloud processing resources. In our system model, W stacks of \acp{GPP} handle baseband processing, following the O-RAN cloudification framework. These \acp{GPP} enable virtualisation due to their programmability and processing power. Workloads are managed by a dispatcher under a global cloud controller near the near-RT-RIC. For CF joint processing, data and control signals for a specific \ac{UE} must be in the same \ac{GPP}, though locally pre-coded signals can be sent to \acp{RU} from different \acp{GPP}, enabled by distributed operations and computational sharing. We utilise the \ac{eCPRI} specification for fronthaul/midhaul transmission, employing a \ac{TWDM-PON} as the fronthaul transport network to meet the high capacity demands of a \ac{CF} network. Each \ac{RU} connects to an  \ac{ONU} assigned to one of multiple wavelengths in the fiber communication, with time-division multiplexing allowing multiple \acp{RU} to share a wavelength. Our power consumption model has two main components: the radio site power, which includes \acp{RU} power $P_{RU,l}$ for each unit ($l = 1, \dots, L$) and \ac{ONU} power $P_{ONU}$, and the Cloud power, calculated using a load-dependent GPP model. The O-Cloud contains an \ac{OLT} with a \ac{WDM MUX} and multiple \acp{LC}, each linked to a \ac{GPP}. Each \ac{LC} serves a single wavelength, ensuring each O-RU’s signals are handled by the \ac{GPP} on the same wavelength. Power savings can be achieved by shutting down unused \acp{GPP} and their corresponding \acp{LC} to minimise the network's active idle power. However, this requires balancing the number of active \acp{GPP} with the \ac{SE} performance of \acp{UE}. 

\begin{figure}
    \centering
\includegraphics[width=0.9\linewidth]{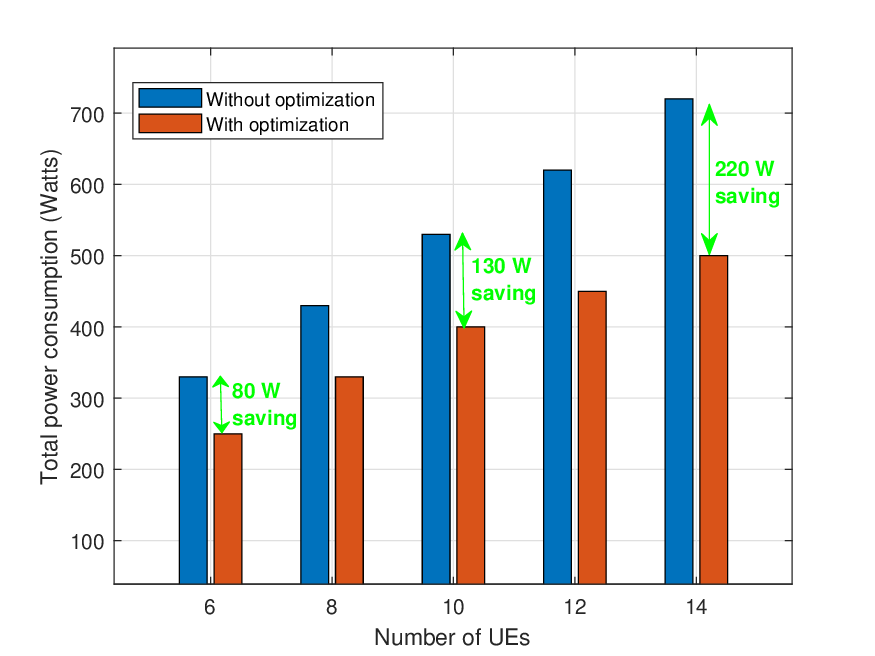}
    \caption{Total power consumption vs. spectral efficiency requirement per UE for $K = 10$ .}
    \label{fig_4}
\end{figure}

\begin{figure}
    \centering
\includegraphics[width=0.9\linewidth]{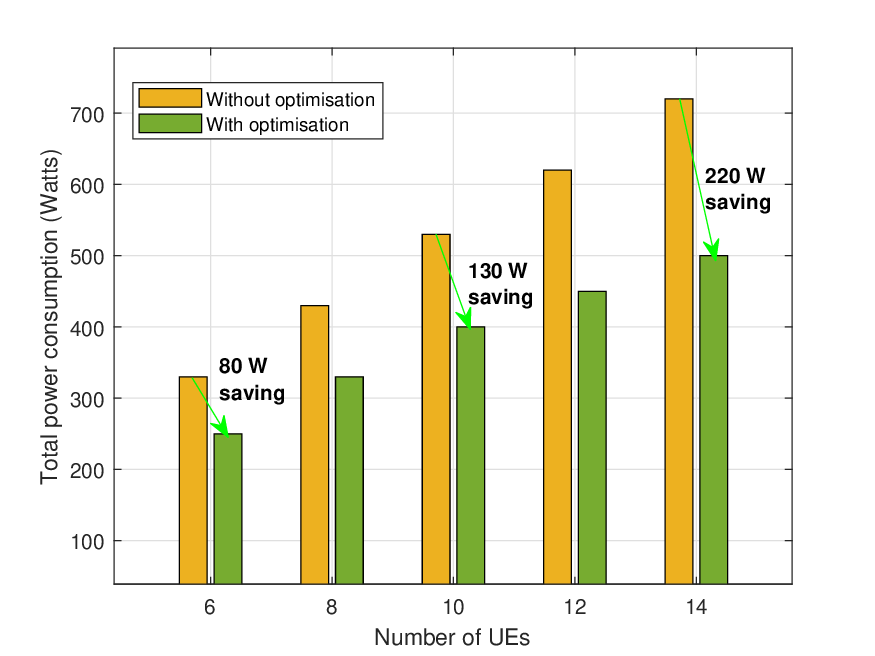}
    \caption{Total power consumption vs. number of UEs for spectral efficienc requirement of 1.5 bit/s/Hz}
    \label{fig_5}
\end{figure}

Figures \ref{fig_4} and \ref{fig_5} demonstrate the impact of optimization on total power consumption in a \ac{CF}-mMIMO architecture. Figure \ref{fig_4} illustrates the relationship between total power consumption and the \ac{SE} requirement per \ac{UE} for a fixed number of UEs ($K = 10$). The results show a significant reduction in power consumption when optimization techniques are applied, achieving up to 190 W of power savings at higher SE requirements. Figure \ref{fig_5} depicts the total power consumption versus the number of UEs for a constant SE requirement of 1.5 bits/s/Hz. Similarly, the optimized configuration consistently demonstrates lower power consumption, with savings reaching up to 220 W as the number of \acp{UE} increases. This underscores the scalability of the optimization techniques within the O-RAN framework, particularly in handling varying user densities while maintaining energy efficiency.

These improvements clearly demonstrate the benefits of implementing a CF-Architecture through the O-RAN framework. The framework's ability to provide flexible optimization of radio, optical fronthaul, and virtualized cloud processing resources is critical. By dynamically managing these resources, O-RAN enables the system to adapt efficiently to fluctuating \ac{SE} and user density requirements, thereby enhancing energy efficiency. The choice of functional split, Option 7.2\footnote{Option 7.2 and Option 8 refer to functional splits within the O-RAN architecture that influence where data processing occurs: Option 7.2 allows for FFT/IFFT and beamforming at the O-RU, reducing bandwidth needs over the fronthaul, whereas Option 8 mandates all processing at the O-DU, necessitating higher data transmission rates.}, plays a pivotal role by reducing the fronthaul data rate compared to Option 8, while still robustly supporting \ac{CF} joint transmission/reception processing and maintaining low O-RU complexity. 
By examining how \ac{CF} architecture, implemented through O-RAN, addresses and reduces energy consumption while enhancing network efficiency, this case study not only supports but also advances our understanding of the potential for sustainable technologies in shaping the future of telecommunications.

\section{Conclusion and Future directions}
In this work, we studied what sustainability means for future networks and presented a number of approaches on how it could be delivered for \ac{6G} and beyond. We argued that Open RAN with its open interfaces and softwarised architecture has the potential to make future mobile networks more sustainable. O-RAN architecture facilitates creation of an ecosystem in which network diversification and continuous and open innovation thrive. In this ecosystem, incremental and radical sustainability related innovations are nurtured to improve \ac{6G}'s sustainability, mainly focusing on \ac{6G}'s energy consumption and e-waste generation. We also highlighted the role of \ac{AI} and how it can help O-RAN to achieve greater sustainability in \ac{6G}. 

It should be noted that with O-RAN's network diversification also comes integration and interoperability challenges which also affect \ac{EE} of O-RAN. Many ongoing research and development activities are focusing on interoperability and integration improvements in O-RAN to improve its sustainability metrics. Since full interoperability has not been achieved yet, activities on this domain will continue to be explored in the future.

 In this work, we introduced sustainability concepts like energy sufficiency for mobile networks. However, currently there are no metrics to measure/benchmark the progress towards achieving them. One direction of future research will be on finding ways to embed these sustainability concepts in \ac{6G} and beyond \acp{KPI}, define metrics for them, and raise public awareness on ways of improving networks' sustainability. 

\begin{acronym} 

\acro{3GPP}{Third Generation Partnership Project}
\acro{3D-InteCom}{Three-Dimensional Integrated Communications}
\acro{5G}{the fifth generation of mobile networks}
\acro{6G}{sixth generation of mobile networks}
\acro{ACO}{Ant Colony Optimization}
\acro{AI}{Artificial Intelligence}
\acro{AR}{Augmented Reality}
\acro{ANN}{Artificial Neural Network}
\acro{AP}{access point}
\acro{BB}{Base Band}
\acro{BBU}{Base Band Unit}
\acro{BER}{Bit Error Rate}
\acro{BS}{Base Station}
\acro{BW}{Bandwidth}
\acro{BIRCH}{Balanced Iterative Reducing and Clustering using Hierarchies}
\acro{BE}{Best-Effort}
\acro{BigCom}{Big Communications}
\acro{CAV}{Connected and Autonomous Vehicle}
\acro{CC}{Chain Composition}
\acro{CF}{Cell-Free}
\acro{C-RAN}{Cloud Radio Access Networks}
\acro{CAPEX}{Capital Expenditure}
\acro{CoMP}{Coordinated Multipoint}
\acro{COTS}{Commercial-Off-The-Shelf}
\acro{CR}{Cognitive Radio}
\acro{CU}{Central Unit}
\acro{COC}{Computation Oriented Communications}
\acro{CAeC}{Contextually Agile eMBB Communications}
\acro{CBQ}{Class-Based Queueing}
\acro{D2D}{Device-to-Device}
\acro{DA}{Digital Avatar}
\acro{DAC}{Digital-to-Analog Converter}
\acro{DAS}{Distributed Antenna Systems}
\acro{DBA}{Dynamic Bandwidth Allocation}
\acro{DNN}{Deep Neural Network}
\acro{DC}{Duty Cycle}
\acro{DyPr}{`Dynamic Prioritization'}
\acro{DL}{Deep Learning}
\acro{DSA}{Dynamic Spectrum Access}
\acro{DT}{Digital Twin}
\acro{DRL}{Deep Reinforcement Learning}
\acro{DQL}{Deep Q Learning}
\acro{DDQL}{Double Deep Q Learning}
\acro{DTs}{Decision Tress}
\acro{DDPG}{Deep Deterministic Policy Gradient}
\acro{DPI}{Deep Packet Inspection}
\acro{$E^2D^2PG$}{Enhanced Exploration Deep Deterministic Policy}
\acro{DU}{Distributed Unit}
\acro{EE}{Energy Efficiency}
\acro{e-waste}{Electronic waste}
\acro{ER}{Erdős-Rényi}
\acro{EUB}{Expected Upper Bound}
\acro{EDuRLLC}{Event Defined uRLLC}
\acro{EVNFP}{Elastic Virtual Network Function Placement}
\acro{eMBB}{enhanced Mobile Broadband}
\acro{eMBB-Plus}{Enhanced Mobile Broadband Plus}
\acro{eCPRI}{evolved CPRI}
\acro{FBMC}{Filterbank Multicarrier}
\acro{FEC}{Forward Error Correction}
\acro{FG} {Forwarding Graph}
\acro{FGE}{FG Embedding}
\acro{FIFO}{First-in-First-out}
\acro{FCFS}{First Come, First Served}
\acro{FFR}{Fractional Frequency Reuse}
\acro{FSO}{Free Space Optics}
\acro{fBm}{Fractional Brownian motion}
\acro{GA}{Genetic Algorithms}
\acro{gNB}{Next Generation Node B}
\acro{GI}{Granularity Index}
\acro{GM}{Gaussian Mixture}
\acro{GPP}{General-Purpose Processor}
\acro{HAP}{High Altitude Platform}
\acro{HD}{High-Demand}
\acro{HL}{Higher Layer}
\acro{HARQ}{Hybrid-Automatic Repeat Request}
\acro{ICT}{Information and Communication Technology}
\acro{IoE}{Internet of Everything}
\acro{IoT}{Internet of Things}
\acro{ILP}{Integer Linear Program}
\acro{IRS}{Intelligent Reflective Surfaces}
\acro{ISAC}{Integrated Sensing and Communication}
\acro{KPI}{Key Performance Indicator}
\acro{KPM}{Key Performance Measurement}
\acro{KNN}{K-Nearest Neighbour}
\acro{LAN}{Local Area Network}
\acro{LAP}{Low Altitude Platform}
\acro{LC}{line card}
\acro{LL}{Lower Layer}
\acro{LLM}{Large Language Model}
\acro{LR}{Logistic Regression}
\acro{LOS}{Line of Sight}
\acro{LTE}{Long Term Evolution}
\acro{LTE-A}{Long Term Evolution Advanced}
\acro{LRD}{Long-Range Dependence}
\acro{LFGL}{Least-First-Greatest-Last}
\acro{MAC}{Medium Access Control}
\acro{MAP}{Medium Altitude Platform}
\acro{MIMO}{Multiple Input Multiple Output}
\acro{ML}{Machine Learning}
\acro{MME}{Mobility Management Entity}
\acro{mmWave}{millimeter Wave}
\acro{MNO}{Mobile Network Operator}
\acro{MR}{Mixed Reality}
\acro{MTC}{Machine Type Communications}
\acro{MILP}{ Mixed-Integer Linear Program}
\acro{MDP}{Markov Decision Process}
\acro{mMTC}{massive Machine Type Communications}
\acro{NAI}{Network Availability Index}
\acro{NASA}{National Aeronautics and Space Administration}
\acro{NAT}{Network Address Translation}
\acro{NHD}{Not-so-High-Demand}
\acro{NN}{Neural Network}
\acro{NF}{Network Function}
\acro{NFP}{Network Flying Platform}
\acro{NTN}{Non-terrestrial networks}
\acro{NFV}{Network Function Virtualization}
\acro{NS}{Network Service}
\acro{OFDM}{Orthogonal Frequency Division Multiplexing}
\acro{OAM}{Orbital Angular Momentum}
\acro{OPEX}{Operational Expenditure}
\acro{OSA}{Opportunistic Spectrum Access}
\acro{O-RAN}{Open Radio Access Network}
\acro{ONU}{optical network unit}
\acro{OLT}{optical line terminal}
\acro{PAM}{Pulse Amplitude Modulation}
\acro{PAPR}{Peak-to-Average Power Ratio}
\acro{PGW}{Packet Gateway}
\acro{PHY}{physical layer}
\acro{PSO}{Particle Swarm Optimization}
\acro{PT}{Physical Twin}
\acro{PU}{Primary User}
\acro{Pr}{Premium}
\acro{QAM}{Quadrature Amplitude Modulation}
\acro{QoE}{Quality of Experience}
\acro{QoS}{Quality of Service}
\acro{QPSK}{Quadrature Phase Shift Keying}
\acro{QL}{Q-Learning}
\acro{RA}{Resource Allocation}
\acro{RAN}{Radio Access Network}
\acro{RF}{Random Forest}
\acro{RN}{Remote Node}
\acro{RRH}{Remote Radio Head}
\acro{RRC}{Radio Resource Control}
\acro{RRU}{Remote Radio Unit}
\acro{RU}{Radio Unit}
\acro{RL}{Reinforcement Learning}
\acro{RR}{Ridge Regression}
\acro{RT}{Real Time}
\acro{RIS}{Reconfigurable Intelligent Surfaces}
\acro{RIC}{RAN Intelligent Controller}
\acro{SBA}{Service Based Architecture}
\acro{SCH}{Scheduling}
\acro{SLAs}{service Level Agreements}
\acro{SU}{Secondary User}
\acro{SCBS}{Small Cell Base Station}
\acro{SDN}{Software Defined Network}
\acro{SFC}{Service Function Chaining}
\acro{SLA}{Service Level Agreement}
\acro{SNR}{Signal-to-Noise Ratio}
\acro{SON}{Self-Organising Network}
\acro{SAR}{Service Acceptance Rate}
\acro{SE}{spectral efficiency}
\acro{SVM}{Support Vector Machine}
\acro{SLFL}{Simple Lazy Facility Location}
\acro{SURLLC}{Secure Ultra-Reliable Low-Latency Communications}
\acro{TDD}{Time Division Duplex}
\acro{TD-LTE}{Time Division LTE}
\acro{TDM}{Time Division Multiplexing}
\acro{TDMA}{Time Division Multiple Access}
\acro{TWDM-PON}{ time- and wavelength-division multiplexed passive optical network}
\acro{TWT}{Threshold Waiting Time}
\acro{THz}{sub-Terahertz} 
\acro{UE}{User Equipment}
\acro{UAV}{Unmanned Aerial Vehicle}
\acro{URLLC}{Ultra-Reliable Low Latency Communications}
\acro{USRP}{Universal Software Radio Platform}
\acro{UCDC}{Unconventional Data Communications }
\acro{UHD}{Ultra-High Definition}
\acro{VL}{Virtual Link}
\acro{VNF}{Virtual Network Function}
\acro{VNF-FG}{VNF-Forwarding Graph}
\acro{VNF-FGE}{VNF-FG Embedding}
\acro{VM}{Virtual Machine}
\acro{VR}{Virtual Reality}
\acro{WFQ}{Weighted Fair Queuing}
\acro{WDM MUX}{WDM multiplexer}
\acro{XAI}{eXplainable Artificial Intelligence}
\acro{XR}{eXtended Reality}
\end{acronym}
\bibliographystyle{IEEEtran}

\bibliography{O-RAN-sustainability-arXiv.bib}

\end{document}